\documentclass[runningheads]{svmult}

\usepackage{makeidx}   
\usepackage{graphicx}  
\usepackage{subeqnar}  
\usepackage{multicol}  
\usepackage{physprbb}  
\makeindex             

\newcommand{\greeksym}[1]{{\usefont{U}{psy}{m}{n}#1}}
\newcommand{\umu}{\mbox{\greeksym{m}}}
\newcommand{\udelta}{\mbox{\greeksym{d}}}
\newcommand{\uDelta}{\mbox{\greeksym{D}}}

\begin{document}
\title*{
TIMMI2 and VLT-ISAAC Spectroscopy of Circumstellar Dust Disks
  -- A Spatially Resolved 3.3\,$\umu$m PAH Feature Around HD\,100546}
\toctitle{
TIMMI2 and VLT-ISAAC Spectroscopy of Circumstellar Dust Disks
  -- A Spatially Resolved 3.3\,$\umu$m PAH Feature Around HD\,100546}
\titlerunning{
TIMMI2 and VLT-ISAAC Spectroscopy of Circumstellar Dust Disks
}

\author{Vincent~C.~Geers\inst{1}
\and Jean-Charles~Augereau\inst{1}
\and Klaus~M.~Pontoppidan\inst{1}
\and Hans~U.~K\"aufl\inst{2}
\and Anne-Marie~Lagrange\inst{3}
\and Ga\"el~Chauvin\inst{3,4}
\and Ewine~F.~van~Dishoeck\inst{1}}

\authorrunning{V.C.~Geers, J.-C. Augereau, K.M. Pontoppidan et al.}

\institute{Leiden Observatory, P.O. Box 9513, 2300 RA Leiden, The Netherlands
\and ESO, Garching, Germany
\and Laboratoire d'Astrophysique de Grenoble, France
\and ESO, Santiago, Chile}

\maketitle              

\begin{abstract}
We present preliminary results of spectroscopic data obtained with the
La Silla 3.6\,m TIMMI2 and the VLT-ISAAC instruments. We have obtained
23 TIMMI2 N-band and 18 Q-band spectra for our total sample of 36
stars in March 2003 to investigate the evolution of the grain
properties in circumstellar disks. Our sample consists of Herbig
stars, T\,Tauri stars and Vega-like stars, some of which have recently
been identified to be members of nearby ($<$100\,pc) stellar
associations.  Silicate and Polycyclic Aromatic Hydrocarbon (PAH)
emission features are detected and the spatial resolution of TIMMI2 allowed to place upper limits on the
spatial extent of the emission.
We also obtained ISAAC L-band spectra of 4 well-studied disks. For the
Herbig Be star HD\,100546, we detect spatially extended PAH emission
at 3.3\,$\umu$m which corresponds to a source with radius 11$\pm$3\,AU
in this line.
\end{abstract}
\section{Introduction}
In the last decade, circumstellar disks at different evolutionary
stages have been intensively studied from optical to millimeter
wavelengths. Low-mass (T~Tauri) and intermediate-mass (Herbig Ae/Be)
young stars have been shown to have young dusty and gaseous
protoplanetary disks for which imaging can provide constraints on the
structure \cite{dar03,man97,wol03}. 
More evolved gas-free dust
disks around main-sequence stars, studied at sub-millimeter and
near-infrared wavelengths, show a ``lumpy'' structure with gaps, holes
and/or asymmetries in the dust distribution, indicative of
gravitational perturbations by stellar companion(s) or unseen
planet(s) \cite{aug01b,aug04,wya03}.  Near- and mid-infrared
spectroscopy of circumstellar disks around Pre Main Sequence (PMS)
stars at different evolutionary stages has shown a wealth of gas and
dust features, enabling detailed studies of dust mineralogy and its
relation with grain processing and disk evolution \cite{mee01}. The
shape of the 10\,$\umu$m feature can serve as a tracer of grain growth
in disks and the presence of crystalline silicates,
not observed in the ISM,
is interpreted as evidence that the dust composition evolves as the
disk is dissipating and large protoplanetary bodies are forming \cite{prz03,boe03}.
\section{The Spatially Resolved ISAAC Spectrum of HD\,100546}
With ISAAC at VLT we obtained in mid-2000 low resolution (R$\sim$600)
L-band
(2.8--4.2\,$\umu$m) spectra of four old PMS stars using a
0.6$''\times$120$''$ slit with a spatial resolution of
0.146$''$/pixel. For HD\,100546 the long slit was aligned with the
major axis of the disk resolved in scattered light
\cite{aug01a,gra01}. The reduction procedure is described in
\cite{pon03}. It includes the removal of bad pixels and jitter
correction before co-adding 2-D spectra, spectrum extraction,
wavelength and flux calibration.
\vspace*{-0.4cm}
\begin{figure}[!ht]
\hbox to \textwidth
{
\parbox{0.65\textwidth}{
   \includegraphics[width=.65\textwidth]{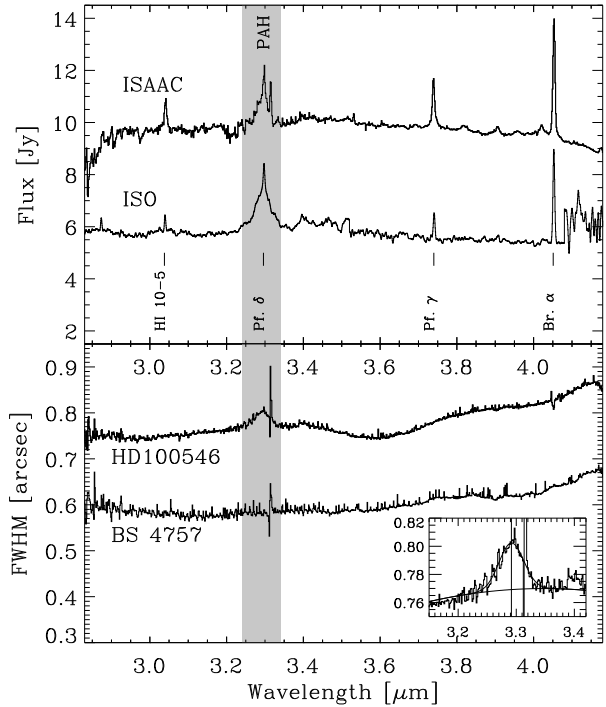}
}
%
\parbox{0.30\textwidth}{
   \caption[]{\label{geers:fig:isaac} ({\bf Upper panel}) ISAAC L-band
spectrum of HD\,100546, scaled by a factor 1.25 (see discussion in the text) to match the ISO
spectrum and then shifted by +4\,Jy for clarity. The grey band highlights the
position of the 3.3 $\umu$m PAH feature. ({\bf Lower panel}) FWHM of the spatial profile, as extracted from 2D spectral image, of HD\,100546 ({\it thick line}) and its corresponding standard BS\,4757 ({\it thin line}). %
The inset shows the FWHM of HD\,100546 only around 3.3\,$\umu$m}
}
}
\end{figure}
\vspace*{-0.5cm}

\noindent The L-band spectrum for HD\,100546 is plotted in the top
panel of Fig.\,\ref{geers:fig:isaac} together with the ISO spectrum.
The two spectra agree reasonably well on absolute flux, although the
ISAAC continuum flux is $\sim$20\% lower than the ISO flux. This
difference cannot be explained by airmass difference between the
science and standard target. The spectrum shows several hydrogen
emission lines, which are stronger in the smaller aperture ISAAC data.
The Pf~$\udelta$ line falls on top of a broad spectral feature at
3.3\,$\umu$m attributed to PAHs \cite{mee01}. Tabulated HI
recombination line ratios from \cite{hum87} were used to estimate and
subtract the contribution of the HI~Pf~$\udelta$ line to the intensity
of the PAH feature. The PAH emission feature to continuum ratios are
similar for both ISAAC and ISO spectra.

In the bottom panel of Fig.\,\ref{geers:fig:isaac}, the measured FWHM of the
spatial profile, assumed to be Gaussian, as a function of wavelength is plotted for HD\,100546
and the standard star BS\,4757, which was used for telluric correction
and flux calibration.
The wavy pattern of the FWHM profiles is currently not understood but
it is similarly observed for the science and standard
stars. Distortion correction of the 2D spectra before FWHM extraction
does not improve the results. The spike at $\sim$3.31\,$\umu$m is caused
by almost complete atmospheric extinction.
When rescaling the measured FWHM of BS\,4757 according to the
difference in measured airmass and DIMM seeing and dividing by a factor
1.05, we find that the rescaled FWHM of BS\,4757 fits reasonably well with the FWHM of
the continuum of HD\,100546. Assuming BS\,4757 is a point
source, we conclude that the continuum of HD\,100546 is unresolved. A
significant bump is observed at 3.3\,$\umu$m with a spatial extent
(FWHM) of 0.22$''\pm0.064'' (3\sigma)$ after correcting for
interpolated FWHM of the continuum. Adopting a Hipparcos distance of
103$^{+7}_{-6}$\,pc \cite{anc97}, this corresponds to a radius of
11$\pm$3\,AU, where the uncertainty is largely due to the noise in
the extracted FWHM. In conclusion we find a significant bump in the FWHM at the wavelength of 3.3\,$\umu$m which indicates that the disk is
spatially resolved in this PAH feature. The 3.3\,$\umu$m PAH carriers
are thus spread over a larger area than the thermalized dust grains
responsible for the adjacent continuum or the ionized gas responsible
for the H lines.
\section{TIMMI2 N-band Spectra of Circumstellar Disks}
\begin{figure}[b]
\begin{center}
\includegraphics[angle=90,width=\columnwidth]{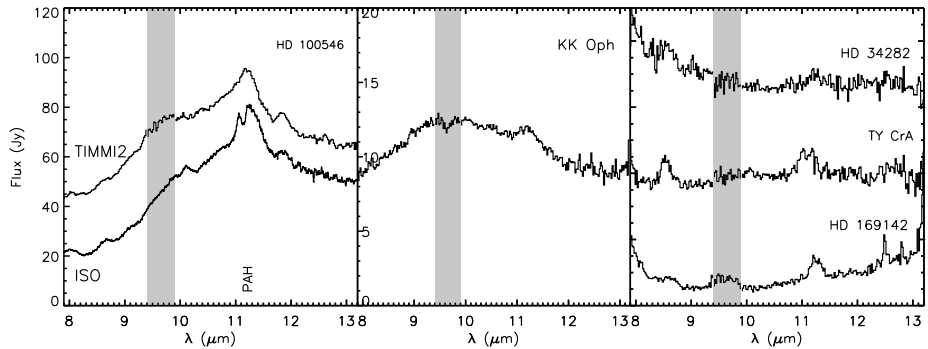}
\end{center}
\caption[]{Selection of TIMMI2 N-band spectra; the grey band indicates
the position of the $\sim$9.6\,$\umu$m atmospheric extinction
band. ({\bf Left panel}) our TIMMI2 spectrum of HD\,100546 is scaled to match the overplotted ISO spectrum and then
shifted +20 Jy for clarity. ({\bf Middle panel}) Spectrum of KK Oph,
normalized to its flux at 12 $\umu$m; showing a broad emission feature
between 9--12\,$\umu$m and a small feature at
$\sim$11.2\,$\umu$m. ({\bf Right panel}) Spectra of HD\,34282 and
TY\,CrA; for comparison our TIMMI2 spectrum of
HD\,169142 is added, for which ISO observations have shown \cite{mee01} that the emission
features at 8.6 and 11--11.2\,$\umu$m can be attributed to the
presence of PAHs}
\label{geers:fig:timmi2}
\end{figure}

With TIMMI2 at the ESO/3.6m telescope we obtained 23 low resolution
N-band (7.5--13.9\,$\umu$m; R$\sim$160; 1.2$''\times$70$''$ slit) and
18 low resolution Q-band (14.5--23.9\,$\umu$m; R$\sim$130;
3.0$''\times$70$''$ slit) spectra for our sample of 36 Herbig Ae/Be
and T Tauri stars with a spatial resolution of 0.45$''$/pixel in
N-band and 0.6$''$/pixel in Q-band. Spectra for targets with IRAS
12\,$\umu$m fluxes as faint as about 0.5--1\,Jy could be observed in
N-band. The reduction procedure is similar to that of the ISAAC
observations. Absolute flux calibration is not performed; for
comparison of features all the spectra are
normalized to their flux at 12\,$\umu$m, except HD\,100546 which is scaled to match the ISO spectrum. Preliminary mid-IR N-band spectra for a selection of our
TIMMI2 sample are shown in Fig. \ref{geers:fig:timmi2}.

For HD\,100546, the shape of our TIMMI2 spectrum agrees reasonably
well with the ISO spectrum but we do observe some interesting
differences: the sharp feature at 11.0\,$\umu$m in the ISO spectrum does not show up in our TIMMI2 spectrum while we do see a more pronounced feature at 11.8\,$\umu$m. The HD\,100546 spectrum has a clear feature at 11.2\,$\umu$m, which can be attributed to blended features from both
forsterite and PAH \cite{bou03}. Using the same approach as for our
ISAAC spectrum, we do not find any evidence of an excess spatial
extent at the 11.2\,$\umu$m feature compared to the extent of the
continuum in our TIMMI2 data, with an upper limit of a radius of 50$\pm$11\,AU.

KK\,Oph, a Herbig Be star, shows a broad emission feature peaking
between 9 and 12\,$\umu$m, which could be attributed to small
amorphous silicate grains, plus additional smaller features like
e.g. that close to 11.2\,$\umu$m, possibly due to crystalline
silicates. TY\,CrA is a triple, perhaps quadruple \cite{cha03},
system and shows clear emission features at 8.5 and 11.1\,$\umu$m of
which the latter feature has, based on ISO-SWS data, been attributed
to PAHs \cite{sie00}. HD\,34282 is another Herbig star for which a
gas disk in Keplerian rotation has recently been resolved at
millimeter wavelengths \cite{pie03}. We detect in its mid-IR spectrum
two features at 8.6 ($\uDelta\lambda=0.28\,\umu$m) and 11.2\,$\umu$m
($\uDelta\lambda=0.20\,\umu$m) again consistent with the presence of
PAHs confirming the tentative detection by ISO.

\section{Future Work with VLT Facilities}
In relation to the work presented here, we plan to use the high spatial and spectral resolution of VISIR to obtain spatially resolved N- and Q-band spectra of young Pre Main Sequence stars with circumstellar disks. This will provide the unique opportunity to simultaneously obtain a clean spectral separation of amorphous and crystalline silicates and distinguish for example the 11.2\,$\umu$m PAH and 11.3\,$\umu$m forsterite features as well as constrain their spatial distribution.
Compared to TIMMI2 on the La Silla 3.6m telescope, VISIR at the 8.2m VLT will allow us to put much stronger constraints on the spatial extent of the 11.2\,$\umu$m PAH feature of HD\,100546 because of the higher spatial resolution.\\\\

\noindent {\it Acknowledgements}:
Astrochemistry in Leiden is supported by a NWO Spinoza grant. J.C.\ Augereau is supported by a fellowship from the European Research Training Network ``The Origin of Planetary Systems'' (PLANETS; contract number HPRN-CT-2002-00308) at Leiden Observatory.

%

\end{document}